\documentclass[a4paper,12pt]{article}
\DeclareMathSizes{12}{12.5}{10}{10}
\usepackage[left=2.3cm,bottom=3cm,right=2.3cm,top=3cm]{geometry}
\usepackage{overpic,youngtab}
\usepackage{subfigure,yfonts}
\usepackage[utf8]{inputenc}
\usepackage[latin,english]{babel}
\usepackage{amsmath}
\usepackage{verbatim}
\usepackage{amssymb}
\usepackage{epsfig}
\usepackage{epstopdf}
\usepackage{scalerel}
\usepackage{float}
\usepackage{fancybox}
\usepackage{color}
\usepackage[colorlinks=true,citecolor=blue,,linktocpage=true,linkcolor=blue,urlcolor=black]{hyperref}
\usepackage{mathabx}

\definecolor{ogreen}{rgb}{0,0.7,0}

\DeclareGraphicsExtensions{.pdf,.png,.jpg,.gif,.jpeg}
\def\be{\begin{equation}}
\def\ee{\end{equation}}
\def\bea#1\eea{\begin{align}#1\end{align}}
\def\pd{\partial}
\def\a{\alpha}
\def\b{\beta}

\def\m{\mu}
\def\n{\nu}

\def\l{\lambda}

\def\r{\rho}

\def\bi{\begin{itemize}}
\def\ei{\end{itemize}}

\def\cL{\mathcal{L}}

\usepackage{authblk}
\usepackage{setspace}

\newcommand{\email}[1]{\href{mailto:#1}{\tt #1}}

\begin{document}


\vspace*{-1cm}
{\flushleft
	{{FTUAM-17-18}}
	\hfill{{IFT-UAM/CSIC-17-090}}}
\vskip 1.5cm
\begin{center}
	{\LARGE\bf A candidate for an UV completion: quadratic gravity in first order formalism\footnote{
	Presented by E.A. in the 18th Lomonosov conference}}\\[3mm]
	\vskip .3cm
	
\end{center}
\vskip 0.5  cm
\begin{center}
	{\Large Enrique Alvarez}, {\Large Jesus Anero},
 {\Large Sergio Gonzalez-Martin} {\large and} {\Large Raquel Santos-Garcia}
\\
\vskip .7cm
{
	Departamento de F\'isica Te\'orica and Instituto de F\'{\i}sica Te\'orica (IFT-UAM/CSIC),\\
	Universidad Aut\'onoma de Madrid, Cantoblanco, 28049, Madrid, Spain\\
	\vskip .1cm

	\vskip .5cm
	\begin{minipage}[l]{.9\textwidth}
		\begin{center} 
			\textit{E-mail:} 
			\email{enrique.alvarez@uam.es},
			\email{jesusanero@gmail.com},
			\email{sergio.gonzalez.martin@uam.es},
			\email{ir.raquel.santos.garcia@gmail.com}
		\end{center}
	\end{minipage}
}
\end{center}
\thispagestyle{empty}

\begin{abstract}
	\noindent
	We consider the most general action for gravity which is quadratic in curvature. In this case first order and second order formalisms are not equivalent. This framework is a good candidate for a unitary and renormalizable theory of the gravitational field; in particular,  there are no propagators falling down faster than $\tfrac{1}{p^2}$. The UV regime is in a  conformal invariant phase; only when Weyl invariance is broken through the coupling to matter can an Einstein-Hilbert term (and its corresponding Planck mass scale) be generated.
\end{abstract}

\section{Introduction.}

It is well-known that general relativity  is not renormalizable
(cf.\cite{Alvarez} and references therein for a general review).
However, quadratic (in curvature) theories are renormalizable,
albeit not unitary \cite{Stelle} -at least in the standard second
order formalism- although they have been widely studied over the years. When considering the Palatini version of the
Einstein-Hilbert lagrangian the connection and the metric are
treated as independent variables and the Levi-Civita connection
appears only when the equations of motion are used.
\par
It is however the case that when more general quadratic in curvature metric-affine actions are
considered  in first order formalism the deterministic relationship between the affine
connection and the Levi-Civita one is lost, even on shell \cite{Borunda}. That
is, the equations of motion do not force the connection to be the
Levi-Civita one.
\par
This is quite interesting because it looks as if  we could  have all the goods of  quadratic lagrangians \cite{Stelle} (mainly renormalizability) without conflicting with  K\"allen-Lehmann's spectral theorem. This justifies our claim of this theory being  a candidate for an ultraviolet (UV) completion of quantum gravity. A preliminary exploration of these ideas has been done in \cite{Alvarez:2017}, to which we refer for a more detailed discussion.

\section{The action principle.}
The action of the theory reads
\be S=\int
d^n
x\sqrt{|g|}\cL\left(g_{\mu\nu},A^{\tau}_{\gamma\epsilon}\right)=\int d^n x\sqrt{|g|}\sum_{I=1}^{I=16}g_I
R^{\mu}_{\quad\nu\rho\sigma}(D_I)_{\mu\mu'}^{\nu\nu'\rho\rho'\sigma\sigma'}R^{\mu'}_{\quad\nu'\rho'\sigma'}\label{Fl}\ee

where the Riemann tensor is defined in terms of an arbitrary (although torsion-free) connection, $\Gamma$, which differs from the Levi-Civita value by a three-index field, $A$:
\be
\Gamma^\m_{\r\l}\equiv \left\{\,^\m_{\r\l}\right\}+A^\m_{\r\l}
\ee
The $D_I$ are straightforward tensors built out of the metric tensor, $g_{\a\b}$. This theory is conformally invariant in $n=4$ dimensions, because under $g_{\m\n}\rightarrow \Omega^2\, g_{\m\n}$ (and $\Gamma$ inert fields)
\be
\cL\rightarrow \Omega^{-4}\,\cL
\ee
It is always possible to choose the background connection as the metric one (that is, the background $A$ field vanishes), in such a way that only quantuum fluctuations in the A field need to be considered

\section{Dynamical generation of the Einstein-Hilbert term.}
The theory so far considered is always in the {\em conformal phase}; it is Weyl invariant. This is the symmetry that prevents the appearance of a cosmological constant on the theory and ensures that all counterterms must be inside our list of quadratic operators. This ultraviolet regime of the theory is then  a candidate for an UV completion of quantum gravity.

This symmetry is not to be found at low energies, however; which means that it must be broken at some scale, which we will relate to Planck's. Once this happens, both a cosmological constant and an Einstein-Hilbert term in the lagrangian are not forbidden anymore.
Several scenarios for this breaking can be proposed (cf. for example, \cite{Salvio});
may be the simplest possibility is through interaction with a minimally coupled  scalar sector
\be
L_s\equiv \sqrt{|g|}\left({1\over 2} g^{\m\n}\pd_\m \phi\pd_\n \phi-V(\phi)\right)
\ee
Quantum corrections will include a term
\be
\Delta L={C\over n-4} R \phi^2
\ee
Were the scalar field to get a nonvanishing vacuum expectation value $\langle\phi\rangle=v$ the counterterm implies an Einstein-Hilbert term
\be
L_{EH}=M^2\sqrt{|g|}R
\ee
Once generated, this term dominates the infrared (IR) phase of the theory.

To conclude, when considering quadratic in the Riemann tensor gravity theories in the first order formalism, quartic propagators never appear. The ensuing theory naively appears to be both renormalizable and unitary.

\par

\section*{Acknowledgments}
E.A. is indebted to Alexander Studenikin for the kind invitation to the Lomonosov conference. Two of us (E.A and S.G-M) are grateful to the LBNL and UC Berkeley  for
hospitality in the initial stages of this project. S. G-M. is also grateful to the University of Southampton (U.K.) for their kind hospitality in the final stages of this work. We are grateful to Bert Janssen, C.P. Mart\'in, Tim R. Morris and Jos Vermaseren for illuminating discussions. Comments by Stanley Deser are always greatly appreciated.
This work has received funding from the European Unions Horizon 2020 research and innovation programme under the Marie Sklodowska-Curie grants agreement No 674896 and No 690575. We also have been partially supported by FPA2012-31880 and FPA2016-78645-P(Spain), COST actions MP1405 (Quantum Structure of Spacetime) and  COST MP1210 (The string theory Universe).  The authors acknowledge the support of the Spanish MINECO {\em Centro de Excelencia Severo Ochoa} Programme under grant  SEV-2012-0249, as well as  by the Spanish Research Agency (Agencia Estatal de Investigación) through the grant IFT Centro de Excelencia Severo Ochoa SEV-2016-0597.

%
%


\end{document}